\documentclass[twocolumn,english,aps,prx,floatfix,amssymb,superscriptaddress,longbibliography]{revtex4}
\usepackage[latin9]{inputenc}
\usepackage{verbatim}
\usepackage{float}
\usepackage{amsmath}
\usepackage{amssymb}
\usepackage{graphicx}
\usepackage{color}
\usepackage{xcolor}
\usepackage{tikz}
\newcommand{\ket}[1]{\ensuremath{\left| #1 \right>}}

\usepackage{bbold}

\usepackage[bookmarks=false,linkcolor=blue,urlcolor=blue,colorlinks,citecolor=blue]{hyperref}
\makeatletter

\newcommand{\be}{\begin{equation}}
\newcommand{\ee}{\end{equation}}
\newcommand{\bea}{\begin{eqnarray}}
\newcommand{\eea}{\end{eqnarray}}

\begin{document}

\title{RG Limit Cycles in BKT Flows $\equiv$ Periodic Real-Time Dynamics in the Current-Current Perturbed $SU(2)_1$ WZW Model}
\author{Parameshwar R. Pasnoori}
\affiliation{Condensed Matter Theory Center, Department of Physics, University of Maryland, College Park, MD 20742, USA}
\affiliation{Mani L. Bhaumik Institute for Theoretical Physics, Department of Physics and Astronomy, \\ University of California, Los Angeles, CA 90095, USA}

\email{parmesh12@g.ucla.edu}

\begin{abstract}
We establish an exact correspondence between renormalization-group (RG) evolution and real-time quantum dynamics in the anisotropic current-current perturbed $SU(2)_1$ Wess-Zumino-Witten (WZW) model. Using its integrable fermionic representation, we construct the exact time-dependent many-body wavefunction by means of the generalized Bethe ansatz and show that periodic boundary conditions lead to quantum Knizhnik-Zamolodchikov equations associated with the XXZ trigonometric $R$-matrix, which corresponds to the quantum affine algebra $\mathcal{U}_q(\widehat{\mathfrak{sl}_2})$ evaluated in its spin-$1/2$ evaluation representation. Consistency of these equations constrains the temporal evolution of the longitudinal and transverse interaction strengths. In the universal regime, these integrability conditions are exactly equivalent to the Berezinskii-Kosterlitz-Thouless RG flow equations upon identifying physical time with the logarithmic RG scale. Consequently, the RG limit cycles of the anisotropic current-current perturbed $SU(2)_1$ WZW model are realized as periodic real-time evolution of its interaction strengths. Our results provide an exact dynamical realization of RG limit cycles and establish a direct connection between cyclic RG flows, quantum integrability, and periodically driven interacting quantum field theories.
\end{abstract}
\maketitle

\section{Introduction}

As a central organizing principle of modern theoretical physics, the renormalization group builds on a profound observation: a system's physical description depends entirely on the scale at which it is observed. Integrating out short-distance degrees of freedom transforms rigid microscopic parameters into dynamic, scale-dependent quantities. This process cleanly separates universal features of a theory from those dictated by microscopic details, profoundly reshaping our understanding of quantum field theory, critical phenomena, statistical mechanics, and many-body systems. 

The renormalization group (RG) emerged from the study of ultraviolet divergences in quantum electrodynamics, where the work of Bethe, Schwinger, Feynman, Tomonaga, Dyson, and others established that divergent contributions could be absorbed into redefinitions of a small number of physical parameters \cite{BetheRG,FeynmannRG,SchwingerRG,TomonagaRG,DysonRG}. The subsequent work of Stueckelberg and Petermann \cite{StueckelbergRG}, and of Gell-Mann and Low \cite{GellmannLowRG}, made explicit that renormalized couplings depend on the scale at which a theory is probed. This scale dependence is encoded in beta functions, whose integral curves define trajectories in the space of couplings. The modern interpretation of these trajectories was developed through Kadanoff's block-spin picture \cite{KadanoffRG} and Wilson's formulation of the RG as a flow through the space of effective theories \cite{WilsonRG1,WilsonRG2,WilsonRG3}. In this framework, changes of scale induce a dynamical evolution of the couplings, and the qualitative behavior of a theory is determined by the resulting RG trajectory \cite{callan,Symanzik}. Much of the conventional discussion of RG flows focuses on fixed points, where the beta functions vanish and the theory becomes scale invariant. The behavior near such points underlies universality, critical scaling, and classifies perturbations as relevant, irrelevant, and marginal. Familiar examples include asymptotic freedom in non-Abelian gauge theories \cite{GrossWilczek,Politzer}, where the ultraviolet flow approaches weak coupling.

Fixed points, however, represent only one possible form of asymptotic RG behavior. As Wilson already emphasized, an RG trajectory need not converge to a stationary point and may instead approach a limit cycle \cite{WilsonRG1,WilsonRG2,WilsonCycle}. In this case, the couplings evolve periodically with the logarithm of the RG scale and return to their original values after a finite RG interval. If $g_i(\ell)$ denote the running couplings, a limit cycle is characterized schematically by
\begin{align}
g_i(\ell+\lambda)=g_i(\ell),
\end{align}
for some finite period $\lambda$. The resulting theory is therefore not invariant under arbitrary continuous rescalings, but rather under the discrete transformation associated with one complete traversal of the cycle. Continuous scale invariance is replaced by discrete scale invariance, and observables can acquire log-periodic dependence on the physical scale. Thus, the limit-cycle behavior is qualitatively distinct from the familiar fixed-point picture. Instead of terminating at a scale-invariant theory, the RG trajectory remains recurrent and winds around a closed orbit in coupling space. The RG evolution therefore retains information about a finite scaling period, producing a hierarchy of physically equivalent scales related by a discrete multiplicative factor. Such recurrent flows provide a particularly interesting generalization of conventional RG dynamics. They raise questions regarding the origin of limit cycles, their manifestation in exactly solvable quantum field theories, and their consequences for physical observables.

An especially instructive realization of this phenomenon occurs in two-dimensional quantum field theory. The Kosterlitz-Thouless class of theories provides a simple setting in which multiple couplings undergo a nontrivial flow, and the geometry of the resulting trajectories can be analyzed explicitly. In a particular region of coupling space, the trajectories cease to approach an ultraviolet or infrared fixed point and instead exhibit cyclic behavior \cite{Bernard1}. The model studied by LeClair, Roman, and Sierra provides an integrable realization of this regime in an anisotropic current-current perturbation of the \textit{level-one $SU(2)$ Wess-Zumino-Witten} ($SU(2)_1$ WZW) model  \cite{LeClair}. Their analysis gave a concrete field-theoretic example in which the RG limit cycle is accompanied by nontrivial spectral and scattering signatures, including a periodicity of the S-matrix in the rapidity variable and a Russian-doll scaling structure in the spectrum.

The existence of an RG trajectory suggests an additional question that is conceptually distinct from the usual interpretation of renormalization. In the conventional RG, $\ell$ is not a physical time coordinate: it labels a sequence of effective descriptions obtained by changing the scale at which the same underlying theory is examined. Can an RG trajectory nevertheless emerge as an actual dynamical evolution in physical time? Remarkably, integrable field theories provide precisely such a possibility.

A connection between RG flow and time-dependent integrability was developed classically by Hoare, Levine, and Tseytlin. They considered integrable two-dimensional field theories in which couplings that are constant in the usual formulation are promoted to functions of worldsheet time. Requiring the resulting time-dependent theory to retain a Lax representation imposes nontrivial constraints on these couplings. A key result is that these constraints coincide with the RG flow equations of the corresponding static theory, with the worldsheet time identified with RG time, up to conventions for the choice of RG scale \cite{Benhoare}. This observation provides a direct bridge between renormalization-group evolution and physical time evolution: the RG trajectory of a static theory can be realized as a trajectory in the parameter space of a genuinely time-dependent integrable system.

This correspondence has subsequently been extended to quantum many-body systems and quantum integrable field theories with time-dependent strengths \cite{pasnoorigrossneveu2,PasnooriRG} within the generalized Bethe-ansatz framework \cite{PasnooriKondo}. In this formulation, the standard Bethe-ansatz problem of solving for eigenstates, typically involving a transfer matrix, is replaced by a dynamical problem involving matrix difference equations. In the class of problems studied, these matrix difference equations take the form of quantum Knizhnik-Zamolodchikov (qKZ) equations. The solution to these equations provides the exact wavefunction \cite{PasnooriGrossNeveu,pasnoorikondo2}. The time dependence of the interaction strengths is not arbitrary: the requirement that the quantum evolution preserve the integrable structure constrains the couplings to follow the corresponding RG trajectories \cite{PasnooriRG}. The resulting framework therefore provides a quantum realization of the classical observation that an RG flow can be interpreted as a dynamical trajectory in time.

The relation between RG flow and time-dependent integrability becomes particularly striking in the presence of limit cycles. If the beta functions generate a closed trajectory in theory space, then the identification of RG time with physical time suggests that the same trajectory should appear as a periodic evolution of the couplings in a time-dependent integrable theory,
Such an interpretation would promote the limit cycle from an abstract structure of theory space to a directly realizable dynamical phenomenon. The purpose of this work is to demonstrate this correspondence explicitly in an exactly solvable quantum field theory.

We focus on the anisotropic current-current perturbed $SU(2)_1$ WZW model considered in \cite{LeClair}. This model is also related to the $\lambda$-deformed $SU(2)_1$ WZW model whose integrability structures have been well studied in the literature \cite{Lacroix,Sfetsos}. The action takes the form
\begin{equation}
S=S_{\rm WZW}+\frac{1}{2\pi}\hspace{-0.5mm}\int \hspace{-0.5mm}d^2x\left[4g_{\perp}\left(J^+_R J^-_L+J^-_R J^+_L\right)-4g_{\parallel}J^z_R J^z_L\right],\label{CCWZW}
\end{equation}
where $J^{\alpha}_R, J^{\alpha}_L$ are right and left moving currents. At one loop, the associated Kosterlitz-Thouless beta functions are
\begin{equation}
\frac{d g_{\parallel}}{d\ell}=-4g_{\perp}^2,\qquad\frac{d g_{\perp}}{d\ell}=-4g_{\perp}g_{\parallel},\label{RGeq}
\end{equation}
with RG invariant
\begin{equation}
Q=g_{\parallel}^2-g_{\perp}^2.\label{WZWccRGinv}
\end{equation}

For $Q>0$, the RG trajectories are associated with the fixed points, and the invariant $Q$ determines the scaling dimension of the perturbing operator. In the bosonized description, these fixed-point flows include the massive sine-Gordon regime where the cosine perturbation is relevant for $0<\sqrt{Q}<\infty$. This regime further separates into repulsive ($0<\sqrt{Q}<1/2$) and attractive ($\sqrt{Q}>1/2$) regimes. In the repulsive regime, the fundamental excitations are solitions and antisolitons which have the same mass but opposite `charge'. In the attractive regime, in addition, there exist bound states of solitons and anti-solitons called breathers. In addition to this there exist massless regime where the cosine perturbation is irrelevant and there also exists a regime described by the sinh-Gordon model. For more details see \cite{LeClair}. 

\begin{center}
\begin{figure}
\includegraphics[width=0.75\columnwidth]{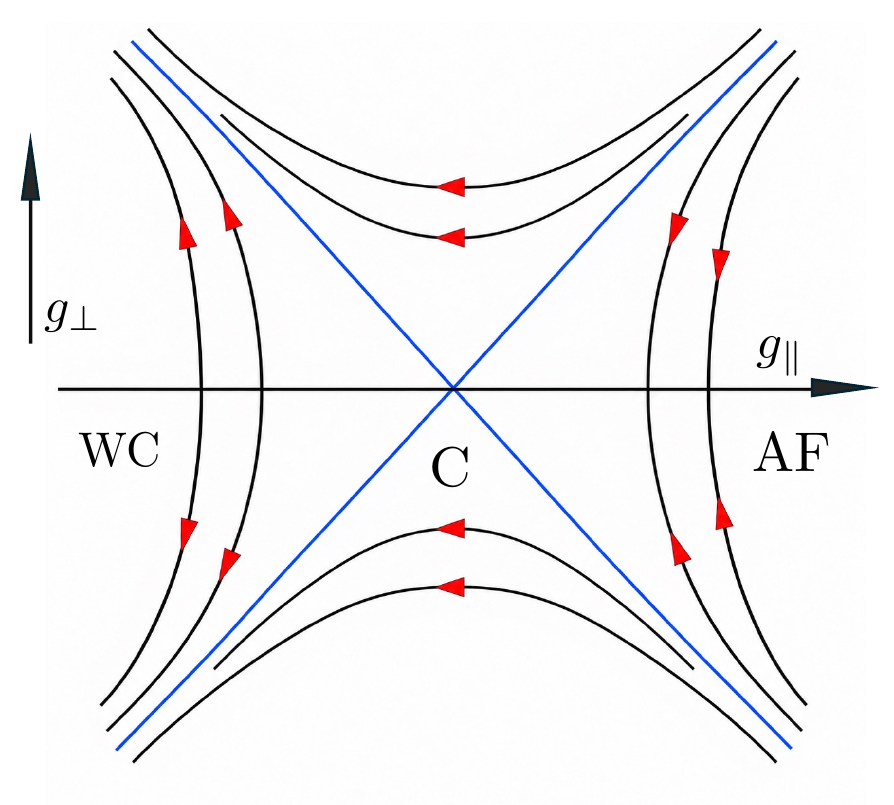}
\caption{The figure depicts the BKT RG flows in various regimes. In the AF (\textit{asymptotic freedom)} regime, which corresponds to $g_{\parallel}>0$, the current-current perturbation is relevant. The system exhibits asymptotic freedom at high energy.  The RG flow lines are monotonic and in the UV they end on the fixed points that lie on the x-axis. In contrast, in the WC (\textit{weak coupling)} regime, the current-current perturbation is  irrelevant. The RG flow lines are monotonic and in the IR, they end on the fixed points that lie on the x-axis. In C (\textit{cross over or cyclic regime}), the RG flow lines are periodic forming limit cycles with a period depending on the RG invariant $Q$. }
\label{fig:picture}
\end{figure}
\end{center}

 By contrast, for $Q<0$ there is no UV or IR fixed point; the flow enters the cyclic regime, where the universal quantity is no longer a fixed-point scaling dimension but the RG period itself. Writing
\begin{equation}
Q=-\frac{h^2}{16},\label{hdef}
\end{equation}
the one-loop solution can be written as
\begin{equation}
g_{\parallel}(\ell)
=-\frac{h}{4}\tan\left[h(\ell-\ell_0)\right],
\end{equation}
which is periodic in RG time, with one-loop period 
\begin{align}\lambda_{\rm 1-loop}=\pi/h. \label{LeClair1loop}\end{align}
LeClair et al. argued that this cyclic behavior persists beyond one loop. Using an all-orders beta-function proposed in \cite{Gerganov}, they found that the period for all orders is given by
\begin{equation}
\lambda_{\rm all-orders}=\frac{2\pi}{h}.\label{LeClairallloop}
\end{equation}
Their analysis further revealed that the same scale periodicity appears in the exact S-matrix \cite{Bernard2},
\begin{equation}
S\left(\beta+\frac{2\pi}{h}\right)=S(\beta),
\end{equation}
and in the associated Russian-doll scaling structure of the spectrum \cite{LeClair,russiandoll}.

 In order to study this corresponding time-dependent problem using the generalized Bethe ansatz, it is useful to formulate the same interacting degrees of freedom in terms of fermions. The appropriate fermionic theory is the $U(1)$-symmetric Thirring model.  The connection between the two descriptions follows from the charge-isospin separation of a doublet of massless Dirac fermions: the charge degrees of freedom form a decoupled $U(1)$ sector, whereas the isospin degrees of freedom are described by the $SU(2)_1$ WZW model. The anisotropic four-fermion interaction of the Thirring model then becomes an anisotropic left-right current-current perturbation of the WZW model. One can promote the interaction strengths of the $U(1)$ Thirring model to explicit functions of time,
\begin{equation}
g_{\parallel}\rightarrow g_{\parallel}(t),
\qquad
g_{\perp}\rightarrow g_{\perp}(t),
\end{equation}
and solve the resulting time-dependent theory using the generalized Bethe ansatz. Schematically, the generalized Bethe-ansatz construction leads to the identification
\begin{equation}
\ell\longrightarrow t,
\end{equation}
so that the RG equations become dynamical equations for the time-dependent couplings,
\begin{equation}
\frac{d g_i(t)}{dt}=\beta_i\big(g(t)\big).
\end{equation}
The cyclic RG trajectory is consequently realized as an actual periodic trajectory in physical time. In the normalization in which RG time and physical time are identified directly, the period of the drive is the same as the period of the RG cycle: In the weak-coupling limit, the temporal cycle maps to the one-loop RG period (\ref{LeClair1loop}) while the exact solution has twice this period and maps to (\ref{LeClairallloop}) thereby reproducing the same factor of two enhancement of the cycle period found beyond one loop. Thus the result provides an explicit realization of the correspondence between the \textit{cyclic RG flow} and \textit{periodic time-dependent coupling strengths}.

Here we stress that the periodic trajectory is not introduced as an externally chosen driving protocol. Rather than starting from a known cyclic RG solution and postulating a time dependence for the Hamiltonian, we begin with the time-dependent quantum field theory and impose the generalized integrability conditions. The resulting equations determine the allowed temporal evolution of the couplings and reproduce the cyclic trajectory previously obtained from the RG analysis. The agreement therefore provides an independent, dynamical derivation of the limit-cycle solution and gives physical meaning to the identification of RG time with time evolution.

The organization of the paper is as follows. In section (\ref{sec:Thirring_WZW}), using non-Abelian bosonization we relate the anisotropic current-current perturbed $SU(2)_1$ WZW model to the $U(1)$ Thirring model. In section (\ref{sec:GBA_Thirring}), we promote the coupling strengths to be time-dependent, and construct the exact generalized Bethe ansatz solution and derive the constraint conditions on the time-dependent couplings. In section (\ref{sec:periodic_driving}) we show that these constraint conditions give rise to periodic time dependent coupling strengths, and we show explicitly that the associated periods are same as the periods of the cyclic RG solution of LeClair et. al,. Finally, in section (\ref{sec:conclusion}) we discuss the implications of interpreting RG limit cycles as periodic integrable dynamics and consider possible extensions to other cyclic RG systems.

\section{Anisotropic Current-Current Perturbation of the $SU(2)_1$ WZW model}
\label{sec:Thirring_WZW}

We introduced the anisotropic current-current perturbed $SU(2)_1$ WZW model and discussed the associated renormalization-group trajectories. To study this model with the time-dependent current-current perturbations, as mentioned above, it is helpful to formulate it in terms of the fermionic field theory which is the $U(1)$ symmetric Thirring model. In this section we establish the relation between these two models using non-Abelian bosonization \cite{WittenNAB}. We begin with a doublet of free massless Dirac fermions and separate the charge and isospin degrees of freedom. The isospin sector is described by the $SU(2)_1$ WZW model, while the charge sector is described by a free $U(1)$ boson. We then introduce anisotropic current-current interactions in the $SU(2)_1$ WZW model and show explicitly that, when expressed in terms of the original fermions, these interactions are precisely those of the $U(1)$-symmetric Thirring model. 

Consider two species of massless Dirac fermions, labeled by an internal isospin index $a=\uparrow,\downarrow$. Writing each Dirac fermion in terms of its right- and left-moving components, $\psi_{Ra}(x)$ and $\psi_{La}(x)$, respectively, the free Hamiltonian is
\begin{equation}
H_0\hspace{-0.7mm}=\hspace{-0.7mm}v_F\hspace{-1mm}\int\hspace{-1mm} dx\hspace{-1mm} \sum_{a=\uparrow,\downarrow}\hspace{-1mm}\left[-i\psi^\dagger_{Ra}(x)\partial_x\psi_{Ra}(x)+i\psi^\dagger_{La}(x)\partial_x\psi_{La}(x)\right],
\label{DiracHamiltonian}
\end{equation}
where $v_F$ is the velocity of the massless fermions, which we will subsequently set $v_F=1$. The free fermionic model possesses both charge and isospin degrees of freedom. Non-Abelian bosonization separates these two sectors. To make this separation explicit, one introduces the chiral charge currents
\begin{align}\nonumber
\rho_R(x)=\sum_a:\psi^\dagger_{Ra}(x)\psi_{Ra}(x):, \\\rho_L(x)=\sum_a:\psi^\dagger_{La}(x)\psi_{La}(x):,
\label{chargecurrents}
\end{align}
and the chiral isospin currents
\begin{align}
J_R^A(x)=\frac{1}{2}\sum_{a,b}:\psi^\dagger_{Ra}(x)\sigma^A_{ab}\psi_{Rb}(x):,
\label{rightspincurrent}
\end{align}
\begin{equation}
J_L^A(x)=\frac{1}{2}\sum_{a,b}:\psi^\dagger_{La}(x)\sigma^A_{ab}\psi_{Lb}(x):,
\label{leftspincurrent}
\end{equation}
where $A=x,y,z$, and $\sigma^A$ are the Pauli matrices acting on the isospin indices $a,b$. The products of fermion fields in Eqs.~(\ref{chargecurrents})-(\ref{leftspincurrent}) are understood in the sense of point-splitting \footnote{The colons correspond to the standard normal-ordering notation with the short-distance singular contribution subtracted}.  The currents $\rho_{R,L}$ generate the chiral $U(1)$ charge algebra, whereas the currents $J^A_{R,L}$ generate the non-Abelian isospin algebra. To display the latter explicitly, let $J^A_{R,n}$ and $J^A_{L,n}$ denote the Fourier modes of the right- and left-moving currents in their respective chiral coordinates. They satisfy
\begin{align}
\left[J^A_{R,m},J^B_{R,n}\right]=i\epsilon^{ABC}J^C_{R,m+n}+\frac{m}{2}\delta^{AB}\delta_{m+n,0},
\label{Kacmoodyright}
\end{align}
and
\begin{align}
\left[J^A_{L,m},J^B_{L,n}\right]=i\epsilon^{ABC}J^C_{L,m+n}+\frac{m}{2}\delta^{AB}\delta_{m+n,0}.
\label{Kacmoodyleft}
\end{align}
Here $\epsilon^{ABC}$ is the completely antisymmetric tensor with $\epsilon^{xyz}=1$, while $\delta^{AB}$ and $\delta_{m+n,0}$ are Kronecker delta symbols. Equations~(\ref{Kacmoodyright}) and (\ref{Kacmoodyleft}) are the affine $SU(2)$ Kac-Moody algebra at level $k=1$,
The right- and left-moving algebras commute,
\begin{align}
\left[J^A_{R,m},J^B_{L,n}\right]=0.
\end{align}
With the normalization of the charge currents in Eq.~(\ref{chargecurrents}), their modes satisfy
\begin{align}\nonumber
\left[\rho_{R,m},\rho_{R,n}\right]=2m\:\delta_{m+n,0},\\
\left[\rho_{L,m},\rho_{L,n}\right]=2m\:\delta_{m+n,0},
\label{KMchargecom}
\end{align}
while the charge and isospin currents commute,
\begin{align}
\left[\rho_{R,m},J^A_{R,n}\right]=\left[\rho_{L,m},J^A_{L,n}\right]=0.
\label{KMchargespincom}
\end{align}
Thus the free fermion theory contains mutually commuting $U(1)$ charge and $SU(2)_1$ isospin current algebras. At the level of local conformal field theories, the corresponding non-Abelian bosonization can therefore be represented schematically as
\begin{align}
U(2)_1\simeq U(1)_{\mathrm{charge}}\times SU(2)_1.
\label{nonabelian_bosonization}
\end{align}
The $U(1)_{\mathrm{charge}}$ sector is equivalently described by a free compact boson, while the isospin sector is the $SU(2)_1$ WZW model. The separation in Eq.~(\ref{nonabelian_bosonization}) is particularly transparent at the level of the Hamiltonian. The free Dirac Hamiltonian in Eq.~(\ref{DiracHamiltonian}) can be represented in terms of the currents defined above, where it takes the following form
\begin{align}
H_0=H_{\mathrm{charge}}+H_{SU(2)_1},
\label{Sugawarafree}
\end{align}
where
\begin{align}
H_{\mathrm{charge}}=\frac{\pi v_F}{2}\int dx\left[:\rho_R^2(x):+:\rho_L^2(x):\right]
\label{Charge_Sugawara}
\end{align}
is the Sugawara Hamiltonian of the $U(1)$ charge sector, and
\begin{align}
H_{SU(2)_1}\hspace{-0.5mm}=\hspace{-0.5mm}\frac{2\pi v_F}{3}\hspace{-1.4mm}\int \hspace{-0.9mm}dx\hspace{-0.4mm}\left[:\hspace{-0.1mm}\boldsymbol{J}_R(x)\hspace{-0.3mm}\cdot\hspace{0.3mm}\boldsymbol{J}_R(x)\hspace{-0.4mm}:\hspace{-0.4mm}+\hspace{-0.4mm}:\hspace{-0.4mm}\boldsymbol{J}_L(x)\hspace{-0.3mm}\cdot\hspace{-0.3mm}\boldsymbol{J}_L(x)\hspace{-0.4mm}:\right]
\label{WZW_Sugawara}
\end{align}
is the Sugawara Hamiltonian of the $SU(2)_1$ WZW sector. Here
\begin{align}
\boldsymbol{J}_R\cdot\boldsymbol{J}_R=\hspace{-1mm}\sum_{A=x,y,z}\hspace{-1mm}J_R^A J_R^A\;\;,\;\;\boldsymbol{J}_L\cdot\boldsymbol{J}_L=\hspace{-1mm}\sum_{A=x,y,z}\hspace{-1mm}J_L^A J_L^A.
\end{align}
From here on we suppress the explicit dependence of the currents on the positions for the sake of notation. The different coefficients in the charge and isospin sectors follow from their respective current algebras: The isospin currents realize an $SU(2)_1$ Kac-Moody algebra, for which the Sugawara coefficient is $(k+h^\vee)^{-1}=1/3$, where the dual Coxeter number of $SU(2)$ is $h^\vee=2$. Equation~(\ref{Sugawarafree}) therefore expresses the charge-isospin separation of the original massless fermionic theory. The central charges provide a simple check of this decomposition. Each massless complex Dirac fermion contributes one unit to the central charge, so that the fermion doublet has $c=2$. The Abelian charge sector has $c=1$, while the $SU(2)_1$ WZW model has
\begin{align}
c_{SU(2)_1}=\frac{3k}{k+2}\bigg|_{k=1}.
\end{align}
The two sectors therefore account for the complete set of degrees of freedom of the free fermionic theory (\ref{DiracHamiltonian}).

Thus the free Dirac Hamiltonian (\ref{DiracHamiltonian}) has separated into a free charge boson (\ref{Charge_Sugawara}) and an $SU(2)_1$ WZW model (\ref{WZW_Sugawara}). We can now perturb the $SU(2)_1$ sector by products of right- and left-moving currents. The most general current-current perturbation that preserves rotations about the $z$ axis while allowing the transverse and longitudinal couplings to differ is
\begin{align}
H_{\mathrm{cc}}=-8\int dx\:\left[g_{\perp}\left(J_R^xJ_L^x+J_R^yJ_L^y\right)+g_{\parallel}J_R^zJ_L^z\right],
\label{CCterm}
\end{align}
where $g_{\perp}$ and $g_{\parallel}$ denote the transverse and longitudinal interaction strengths respectively. It is also useful to introduce the raising and lowering currents
\begin{align}
J_{R,L}^{\pm}=J_{R,L}^{x}\pm iJ_{R,L}^{y}.
\end{align}
Using
\begin{equation}
J_R^xJ_L^x+J_R^yJ_L^y=\frac{1}{2}\left(J_R^+J_L^-+J_R^-J_L^+\right),
\end{equation}
the perturbation may equivalently be written as
\begin{align}
H_{\mathrm{cc}}=-4g_{\perp}\int dx \left(J_R^+J_L^-+J_R^-J_L^+\right)-8g_{\parallel}\int dx J_R^zJ_L^z.
\end{align}
The complete bosonized Hamiltonian is therefore
\begin{equation}
H=H_{\mathrm{charge}}+H_{SU(2)_1}+H_{\mathrm{cc}}.
\label{CompleteBoseHamiltonian}
\end{equation}
The first term in Eq.~(\ref{CompleteBoseHamiltonian}) is the free $U(1)$ charge boson, the second is the $SU(2)_1$ WZW Hamiltonian, and the third is the anisotropic current-current perturbation responsible for the nontrivial RG flow.

We now express the perturbation in terms of the original fermions. From Eqs.~(\ref{rightspincurrent}) and (\ref{leftspincurrent}),
\begin{equation}\label{JRJLfermion}
J_R^A J_L^A\hspace{-0.5mm}=\hspace{-0.5mm}\frac{1}{4}\hspace{-1mm}\sum_{a,b,c,d}\hspace{-1.4mm}\Big[\psi^\dagger_{Ra}(x)\sigma^A_{ab}\psi_{Rb}(x)\Big]\hspace{-0.5mm}\Big[\psi^\dagger_{Lc}(x)\sigma^A_{cd}\psi_{Ld}(x)\Big].
\end{equation}
Substituting Eq.~(\ref{JRJLfermion}) into Eq.~(\ref{CCterm}) one obtains

\begin{align}
H_{\mathrm{cc}}=2\int dx\hspace{-0.5mm}\sum_{a,b,c,d}\hspace{-1mm}\mathcal{A}_{ab,cd}\;\psi^{\dagger}_{Ra}(x)\psi^{\dagger}_{Lc}(x)\psi_{Rb}(x)\psi_{Ld}(x),
\label{Thirring_interaction}
\end{align}
where
\begin{equation}\label{intanisotropic}
{\cal A}_{ab,cd}=g_{\perp}\left(\sigma^x_{ab}\sigma^x_{cd}+\sigma^y_{ab}\sigma^y_{cd}\right)+g_{\parallel}\sigma^z_{ab}\sigma^z_{cd}.
\end{equation}
The current-current perturbation of the $SU(2)_1$ WZW sector is therefore exactly the anisotropic four-fermion interaction. Finally, combining the original free-fermion Hamiltonian with Eq.~(\ref{Thirring_interaction}), Eq.~(\ref{CompleteBoseHamiltonian}) can be written entirely in terms of fermions 
\begin{align}\nonumber
H\hspace{-0.5mm}=&v_F\hspace{-1mm}\int\hspace{-0.7mm} dx\hspace{-1mm}\sum_{a=\uparrow,\downarrow}\hspace{-0.8mm}\left[-i \psi^\dagger_{Ra}(x)\partial_x\psi_{Ra}(x)+i \psi^\dagger_{La}(x)\partial_x\psi_{La}(x)\right]\\
&+2\int dx\hspace{-0.5mm}\sum_{a,b,c,d}{\cal A}_{ab,cd}\; \psi^\dagger_{Ra}(x)\psi^\dagger_{Lc}(x)\psi_{Rb}(x)\psi_{Ld}(x).
\label{Thirringfinalform}
\end{align}
Eq.~(\ref{Thirringfinalform}) is precisely the Hamiltonian of the $U(1)$-symmetric Thirring model. When $g_{\parallel}=g_{\perp}$, the interaction is isotropic and the global isospin symmetry is enlarged to $SU(2)$, in which case the model is called the chiral invariant $SU(2)$ Gross-Neveu model \cite{AndreiLowenstein79,DestriLowenstein}, sometimes refereed to as the $SU(2)$ Thirring model. For $g_{\parallel}\neq g_{\perp}$, only rotations about the $z$ axis remain, giving the $U(1)$ isospin symmetry of the anisotropic theory.

This separation is also manifest in the exact Bethe-ansatz solution of the $U(1)$-symmetric Thirring model \cite{Japaridze}. The charge and isospin degrees of freedom decouple: The eigenstates are characterized by the momenta of the original fermions which are expressed in terms of the charge quantum numbers and a set of rapidities describing the collective isospin degrees of freedom. The charge quantum numbers are pure integers and can be chosen freely reflecting the absence of interaction of the charge degrees of freedom, while the rapidities describing the isospin degrees of freedom are constrained by the coupled transcendental equations called the Bethe equations. The charge-isospin separation observed in the exact solution is therefore the Bethe-ansatz counterpart of the decomposition into $U(1)_{\mathrm{charge}}$ and $SU(2)_1$ sectors described above. The excitations in the charge sector are always massless whereas the four fermion interaction results in the dynamical generation of the mass scale in the spin sector in certain regimes, which we shall discuss below.

The equivalence between the two descriptions is particularly useful for understanding their renormalization-group structure. Since the interacting sector contains the same longitudinal and transverse current-current operators, the weak-coupling RG equations of the $U(1)$-symmetric Thirring model have the same Kosterlitz-Thouless structure as those of the anisotropic current-current perturbed $SU(2)_1$ WZW model. In particular, the RG trajectories are characterized by the same RG invariant (\ref{WZWccRGinv}). It will be helpful to introduce the RG invariant parameter $u$ and the parameter $f$ which flows along the RG that are commonly used in the Bethe ansatz description of the $U(1)$ Thirring model:

\begin{align}
\cos u& =\frac{\cos g_{\parallel}}{\cos g_{\perp}}.\label{RGinvu}\\ \frac{\sin u}{\tanh f}&=\frac{\sin g_{\parallel}}{\cos g_{\perp}}.\label{paramf}
\end{align}
Expanding the above equation (\ref{RGinvu}) to the second order and using (\ref{WZWccRGinv}), one obtains 
\begin{align}u^2\equiv Q.
\end{align}
The crossover regime of the $U(1)$-symmetric Thirring model, characterized by
\begin{equation}
|g_{\perp}|>|g_{\parallel}|,
\end{equation}
therefore corresponds to $Q<0$ or equivalently $u$ being purely imaginary. This is precisely the region of coupling space associated with the cyclic RG trajectories discussed in the previous section. The advantage of the fermionic formulation is that the interacting isospin sector can be treated directly using the Bethe ansatz. This allows us to approach the cyclic RG problem from a different direction. Rather than starting from the RG equations of the perturbed WZW model, we begin with the exact $U(1)$-symmetric Thirring model and subsequently promote its longitudinal and transverse couplings to time-dependent functions. The generalized Bethe ansatz then determines the conditions under which the resulting time-dependent theory remains integrable. As we demonstrate below, these conditions generate the same functional evolution of the couplings that appears in the cyclic RG trajectory of the anisotropic current-current perturbed $SU(2)_1$ WZW theory, with the logarithmic RG parameter replaced by physical time.

Before we proceed further, let us briefly summarize different regimes corresponding to the $U(1)$ Thirring model illustrated in Fig: (\ref{fig:picture}). In the region $|g_{\perp}|<|g_{\parallel}|$, $g_{\parallel}>0$, referred to as $\text{AF}$ regime, which is an abbreviation for \textit{asymptotic freedom regime}, the current-current perturbation is relevant, and the RG flow lines are monotonic and end on the line of fixed points. In this region, the model exhibits a mass gap which is dynamically generated due to the four fermion interaction

\begin{align}m=\Lambda \:\text{arctan}\left(\frac{1}{\sinh(\pi f/2u)}\right)\longrightarrow 2\Lambda e^{-\pi f/2u}.\label{mass}
\end{align}
One takes the scaling limit where the cutoff $\Lambda\rightarrow\infty$ while keeping the physical mass $m$ fixed. This results in the flow of the parameter $f\rightarrow\infty$. The flow of the coupling strengths $g_{\parallel}$ and $g_{\perp}$ along the RG can then be determined through the relations (\ref{RGinvu}), (\ref{paramf}) and noting that the parameter $u$ is RG invariant. One essentially finds that $g_{\perp}$ decreases and goes to zero at the UV fixed points. Thus the model is asymptotically free at high energies. The AF region can be further split into two sub regions depending on the value of the parameter $u$: For $0<u<\pi/2$, the interactions are repulsive and the excitations above the ground state consist of spinons. Note that through Abelian bosonization, the $U(1)$ Thirring model can be shown to be equivalent to the compact bososn CFT which describe the charge degrees of freedom, plus the sine-Gordon model which describes the spin degrees of freedom. The spinons of the $U(1)$ Thirring model are just the solitons of the sine-Gordon model associated with the spin degrees of freedom. For $\pi/2<u<\pi$, the interactions are attractive and in addition to the solitions, the model exhibits bound states of solitons and anti-solitons called breathers. Here, the solitons and anti-solitons are just the spinons with opposite spin orientations. Thus the breathers have zero spin, or equivalently zero charge, in the sine-Gordon language. The breathers also have a finite mass, which decreases as one moves towards the semi-classical limit $u\rightarrow \pi$. The structure of the excitations above the ground state in the attractive regime is mathematically very pleasing: At every rational value of the parameter $u$, a new breather bound state emerges and as one moves towards the semi-classical limit, more breathers emerge and simultaneously their masses decrease. This picture is even more profound in the case of open boundary conditions whose exact Bethe ansatz solution was found in \cite{pasnoori2025duality}. In contrast, for $g_{\parallel}<0$, $|g_{\perp}|<|g_{\parallel}|$, which is referred to as the WC regime, abbreviation for  \textit{weak coupling regime}, the current-current perturbation is irrelevant, and thus the spin sector is massless. For $|g_{\perp}|>|g_{\parallel}|$, which is referred to as C regime, traditional abbreviation for \textit{cross over regime}, or in the current context \textit{cyclic regime}, where the RG flow lines are cyclic, the model is shown to exhibit a mass gap which depends on the value of $g_{\parallel}$. The mass gap is shown to exponentially vanish as one moves from the separatices adjoining the AF regime towards the separatrices adjoining the WC regime. For more details we refer the reader to \cite{Japaridze}.

\section{Generalized Bethe ansatz solution}
\label{sec:GBA_Thirring}

Having established the relation between the anisotropic current-current perturbed $SU(2)_1$ WZW model and the $U(1)$-symmetric Thirring model, we now construct the exact many-body wavefunction of the latter when the interaction strengths are explicitly time dependent. We consider periodic boundary conditions and apply the generalized Bethe ansatz developed for quantum field theories with time-dependent interactions \cite{PasnooriGrossNeveu}. The essential ingredients are the two-particle scattering (S)- matrices and the consistency conditions obtained by transporting particles around the system and applying periodic boundary conditions. Ref.~\cite{PasnooriGrossNeveu} considered the chiral invariant $SU(2)$ Gross-Neveu model, which corresponds to the separatrix between the AF regime and the C regime. It was shown that the two-particle scattering matrices take the form of the XXX $R$-matrix and the consistency conditions lead to the corresponding Yang-Baxter equations which constrain the allowed time-dependent strengths. By transporting a particle around the system and applying periodic boundary conditions results in the qKZ equations associated with the XXX $R$-matrix, whose solutions provide the exact wavefunction. In the current case, breaking the $SU(2)$ symmetry to $U(1)$ replaces the rational XXX $R$-matrix with its $q$-deformation, the trigonometric XXZ $R$-matrix, which arises from the spin-1/2 evaluation representation of the universal $R$-matrix for the quantum affine algebra $U_q(\widehat{\mathfrak{sl}}_2)$.  As we shall show, the resulting consistency conditions constrain the temporal evolution of the two interaction strengths $g_{\parallel}(t)$ and $g_{\perp}(t)$ and the corresponding qKZ equations involve the XXZ $R$-matrix. More importantly, we show that in the universal weak-coupling regime the constraints on the interaction strengths reproduce the renormalization-group equations of the perturbed $SU(2)_1$ WZW model (\ref{CompleteBoseHamiltonian}), with the logarithmic RG scale replaced by physical time. 

\subsection{Hamiltonian and the S-matrix}

We consider the system on a ring of circumference $L$. The Hamiltonian is
\begin{equation}
H_{\mathrm{TH}}(t)=\int_0^L dx\:{\cal H}_{\mathrm{TH}}(x,t),
\label{eq:TH_Hamiltonian}
\end{equation}
where
the Hamiltonian density $\mathcal{H}_{\mathrm{TH}}(x,t)$ is given by (\ref{Thirringfinalform}), (\ref{intanisotropic}) with 
 \begin{align} g_{\parallel}\longrightarrow g_{\parallel}(t), \;\; g_{\perp}\longrightarrow g_{\perp}(t).
 \end{align}
We apply periodic boundary conditions on the fermion fields:
\begin{equation}
\psi_{Ra}(x+L)=\psi_{Ra}(x),
\qquad
\psi_{La}(x+L)=\psi_{La}(x).
\label{PBC}
\end{equation}
Since the bulk interaction only scatters a right mover from a left mover without changing their chiralities, the number of left and right moving particles are separately conserved, thus one can construct the many-body states with $N$ particles in a sector containing $N_R$ right movers and $N_L$ left movers, with $N=N_R+N_L$
\begin{equation}
i\partial_t\left|N_L,N_R;t\right\rangle=H_{\mathrm{TH}}(t)\left|N_L,N_R;t\right\rangle .
\label{SE}
\end{equation}
The one particle sector is trivial as there exist no interactions, and so is the two particle sector containing two left movers or two right movers. To understand the structure of the solution, it is sufficient to begin with the two-particle sector containing one right mover and one left mover. We write
\begin{equation}
|1,1;t\rangle\hspace{-0.5mm}=\hspace{-0.5mm}\sum_{a,b}\hspace{-0.7mm}\prod_{i=1,2}\int_0^L\hspace{-2.2mm}dx_i\psi^\dagger_{Ra}(x_1)\psi^\dagger_{Lb}(x_2)\mathcal{A}F^{RL}_{ab}(x_1,x_2,t)|0\rangle.
\end{equation}
Here $\mathcal{A}$ is the anti-symmetrization symbol. Substituting this into Eq.~(\ref{SE}), one obtains
\begin{align}\nonumber
&\big[-i\left(\partial_t+\partial_{x_1}-\partial_{x_2}\right)I_{ab,cd}\\&+2\delta(x_1-x_2){\cal A}_{ab,cd}(t)\big]\mathcal{A}F^{RL}_{cd}(x_1,x_2,t)=0,
\label{2pseth}
\end{align}
where repeated isospin indices are summed.  Away from the point of scattering $x_1=x_2$, Eq.~(\ref{2pseth}) is a first-order differential equation describing free fermions. Since the particles interact when they scatter, one needs to distinguish between different orderings of the particles in the configuration space. Thus we take the following ansatz
\begin{align}
F^{RL}_{ab}(x_1,x_2,t)=f^{RL,12}_{ab}(z_1,\bar z_2)\theta(x_2-x_1)
\nonumber\\\
+f^{RL,21}_{ab}(z_1,\bar z_2)\theta(x_1-x_2),
\label{2pansatzTH}
\end{align}
where $\theta(x)$ is the Heaviside step function \footnote{ Unlike the case of $SU(2)$ Gross-Neveu model \cite{PasnooriGrossNeveu}, here we do not use any particular regularization of the Heaviside function}. Here we have introduced the coordinates $z_1=x_1-t, \bar z_2=x_2+t$. The first superscript in the amplitudes on the right side of the above equation denotes the chirality of the particles, whereas the second superscripts specify the ordering of the two particles in configuration space. Using the above ansatz (\ref{2pansatzTH}) in (\ref{2pseth}) and requiring that the delta-function contribution generated when derivatives act on the step functions in Eq.~(\ref{2pansatzTH}) must cancel the interaction term in Eq.~(\ref{2pseth}) gives rise to the following relation between the amplitudes
\begin{equation}
f^{RL,21}(z_1,\bar z_2)=S_{12}(z_1,\bar z_2)f^{RL,12}(z_1,\bar z_2),
\label{2prel}
\end{equation}
where $S_{12}(z_1,\bar{z}_2)$ acts on the tensor product of the isospin spaces of particles $1$ and $2$ and takes the following form
\begin{equation}
S_{12}(z_1,\bar z_2)=e^{i\phi_{12}(z_1,\bar z_2)}R_{12}\left(f\left(\frac{\bar z_2-z_1}{2}\right)\right),\label{2psmat}
\end{equation}
where $e^{i\phi_{12}}$ is an overall scalar scattering phase and $f(t)$, to be defined below, is the spectral parameter associated with the time at which the right- and left-moving particles cross. Since
\begin{equation}
\frac{\bar z_2-z_1}{2}=t
\end{equation}
at the scattering point $x_1=x_2$, the argument of $f(t)$ is precisely the physical time of scattering. For more details we refer the refer to the supplementary material of \cite{PasnooriKondo}. For $|g_{\parallel}|>|g_{\perp}|$, $R_{12}(\lambda)$ takes the form of the XXZ $R$-matrix with hyperbolic functions:
\begin{align}
R_{12}(\lambda)
&=\begin{pmatrix}\nonumber
1 & 0 & 0 & 0\\
0 &
\dfrac{\sinh\lambda}{\sinh(\lambda+iu(t))}
&
\dfrac{i\sin u(t)}{\sinh(\lambda+iu(t))}
&0\\
0&
\dfrac{i\sin u(t)}{\sinh(\lambda+iu(t))}
&
\dfrac{\sinh\lambda}{\sinh(\lambda+iu(t))}
&0\\
0&0&0&1
\end{pmatrix}.\\ & \hspace{12mm} (\text{in the case} \;\;\;|g_{\parallel}|>|g_{\perp}|).
\label{XXZRmat}
\end{align}
Here, the parameter $u(t)$ is the crossing parameter. The parameters $f(t)$ and $u(t)$ are related to the time-dependent coupling strengths $g_{\parallel}(t)$ and $g_{\perp}(t)$ through the relations

\begin{align}
f(t)=\operatorname{arccoth}
\left[\frac{\sin^2 g_{\parallel}(t)}{\sin\left[g_{\parallel}(t)-g_{\perp}(t)\right]\sin\left[g_{\parallel}(t)+g_{\perp}(t)\right]}\right]^{1/2},
\label{f(t)def}
\end{align}
\begin{align}
\cos u(t)=\frac{\cos g_{\parallel}(t)}
{\cos g_{\perp}(t)}.\label{u(t)def}
\end{align}

\textit{In the case of interest where $|g_{\parallel}|<|g_{\perp}|$, the S-matrix is given by the transformation $f(t)\rightarrow if(t)$ and $u\rightarrow iu(t)$ in the above equations (\ref{XXZRmat}), (\ref{f(t)def}) and (\ref{u(t)def}), where the S-matrix corresponds to the XXZ $R$-matrix with trigonometric functions. }

 Thus we have constructed the exact wavefunction in the non trivial sector corresponding to the case of two particles $N=2$.  The two amplitudes in the wavefunction (\ref{2pansatzTH}) are related to each other through the relation (\ref{2prel}), thus if one amplitude is know, the other amplitude and thus the complete form of the exact wavefunction can be found. Thus, we are left to determine the remaining amplitude. To achieve this, one can apply periodic boundary conditions (\ref{PBC}) which give rise to the matrix difference equation similar to the $SU(2)$ case which imposes constraints on the amplitudes
\begin{align}
f^{RL,12}(z_1,\bar z_2+L)=S_{12}(z_1,\bar z_2)f^{RL,12}(z_1,\bar z_2).\label{diffeq2p}
\end{align}
One can solve the above matrix difference equation and thus find the exact form of the two particle wavefunction. Note that just as in the case of $SU(2)$ \cite{PasnooriGrossNeveu}, no restrictions are imposed on the time-dependent strengths in the two particle sector. However, this is no longer the case in the $N\geq 3$ particle sector. As we shall see, integrability imposes constraints on the time-dependent coupling strengths.

\subsection{Yang-Baxter equations, consistency of the wavefunction and the qKZ equations}

We now generalize the construction to an arbitrary number of particles $N$. We have

\begin{align}\nonumber
\ket{N_L,N_R;t}=\prod_{j=N_{L}+1}^{N}\prod_{k=1}^{N_L}\int_{0}^L\hspace{-2.5mm}dx_j\hspace{-1mm}\int_{0}^L\hspace{-2.5mm}dx_k \; \psi^{\dagger}_{R\sigma_j}(x_j)\psi^{\dagger}_{L\sigma_k}(x_k)\\\times\mathcal{A}F^{1...N,\{\chi_i\}}_{\sigma_1...\sigma_N}(x_1,...,x_N,t)\ket{0}.\label{npwf1}
\end{align}
Here, $\sigma_1...\sigma_N\equiv\{\sigma_i\}$ denote the spin and $\{\chi_i\}\equiv \chi_1...\chi_N$ denotes the chiralities of the particles. Using this in the Schrodinger equation (\ref{SE}), we obtain the following $N$ particle Schrodinger equation 

\begin{align}\nonumber
 &-i(\partial_t+\hspace{-3mm}\sum_{j=N_L+1}^{N}\hspace{-3mm}\partial_{x_j}\hspace{-1mm}-\hspace{-1mm}\sum_{k=1}^{N_L}\partial_{x_k})\mathcal{A}F^{1...N,\{\chi_i\}}_{\sigma_1...\sigma_N}(x_1,...,x_N,t)+\\&\sum_{\substack{j=N_L+1\\k=1}}^{\substack{j=N\\k=N_L}}\hspace{-3mm}\delta(x_j-x_k)A_{\sigma_j\sigma'_j,\sigma_k\sigma'_k}(t)\mathcal{A}F^{1...N,\{\chi_i\}}_{\sigma_1..\sigma'_j\sigma'_k..\sigma_N}\hspace{-0.4mm}(x_1,...,x_N,t)\hspace{-0.7mm}=\hspace{-0.5mm}0.\label{senp}
\end{align}
The ansatz for $F^{1...N,\{\chi_i\}}_{\sigma_1...\sigma_N}(x_1,...,x_N)$ takes the following form

\begin{align} 
F^{1..N,\{\chi_i\}}_{\sigma_1...\sigma_N}\hspace{-0.3mm}(x_1,..,x_N,t)
\hspace{-1mm}= \hspace{-1.3mm}\sum_Q \theta(\{x_{Q(j)}\})  f^{Q,\{\chi_i\}}_{\sigma_1...\sigma_N}(\bar{z}_1,..,z_N).\label{npwf2}
\end{align}
In this wavefunction, without losing generality, we have chosen the particles $i=1,...,N_L$ to be left movers, and $i=N_L+1,...,N$ to be right movers. In the above expression, $Q$ denotes a permutation of the position orderings of particles and  $\theta(\{x_{Q(j)}\})$ is the Heaviside function that vanishes unless $x_{Q(1)} < \dots < x_{Q(N)}$. Here $f^{Q,\{\chi_i\}}_{\sigma_1...\sigma_N} (\bar{z}_1,...,z_N)$ is the amplitude corresponding to the ordering of the particles denoted by $Q$. The amplitudes that differ by the ordering of the particles with different chiralities are related by the S-matrix (\ref{2psmat}), just as in the two particle case
\begin{align}
f^{...kj...,\{\chi_i\}} (\bar{z}_1,...,z_N)=S^{jk}(z_j,\bar{z}_k)f^{...jk...,\{\chi_i\}} (\bar{z}_1,...,z_N).\label{rel1N}
\end{align}
Here $\chi_j=+,\chi_k=-$ and $``..."$ in the first superscript on both side of the above equation corresponds to any specific ordering of the rest of the particles, which is the same on both side of the equation. We shall use this notation throughout the manuscript without explicitly stating it. Unlike the particles with opposite chiralities, the Hamiltonian does not directly constrain the amplitudes corresponding to the exchange of the particles with the same chiralities, because their trajectories never cross physically. Integrability fixes the relation between these amplitudes by requiring consistency when these particles scatter with particles of opposite chirality. The corresponding matrices are again XXZ $R$-matrices, now depending on differences of spectral parameters. For the case of two right moving particles, we have
\begin{align}
f^{...kj...,\{\chi_i\}}(\bar{z}_1,...,z_N)=S^{jk}(z_j,z_k)f^{...jk...,\{\chi_i\}}(\bar{z}_1,...,z_N),\label{rel3N}
\end{align}
where $\chi_{j,k}=+$.  Here $S^{jk}(\lambda)$ takes the same form as (\ref{2psmat}). The case of two left moving particles is similar to the above case and is obtained by the transformation $z_j\rightarrow\bar{z}_j, z_k\rightarrow\bar{z}_k$ in the above equation.  Note that, the form of the above S-matrices suggests a simple way of viewing the above relations: When a right moving particle $j$ located at $x_j$ is moved past another right moving particle $k$ at position $x_k$, the argument of the S-matrix is simply proportional to the distance the particle $j$ has to travel to move past particle $k$, which is $x_k-x_j=z_k-z_j$. The same applies to the two left moving particles as well.

The consistency of different sequences of pairwise exchanges requires that the S-matrices satisfy the Yang-Baxter equation. For one right moving particle $i$ and two left moving particles $j$ and $k$, we have
\begin{align}\nonumber
S^{ij}(z_i,\bar{z}_j)S^{ik}(z_i,\bar{z}_k)S^{jk}(\bar{z}_j,\bar{z}_k)=\\S^{jk}(\bar{z}_j,\bar{z}_k)S^{ik}(z_i,\bar{z}_k)S^{ij}(z_i,\bar{z}_j).\label{YBn1}\end{align}
This condition requires that the crossing parameter $u(t)$ is time independent 
\begin{align} u(t)\equiv \text{constant},\label{constraintu(t)}
\end{align}
thus comparing the above equation with the case of constant strength (\ref{RGinvu}), we find that the RG invariant of the static model turns into a dynamical invariant in the time-dependent model \cite{pasnooriBGBA}. In contrast to the crossing parameter, the spectral parameter $f(t)$ is allowed to evolve with time. Because the XXZ $R$-matrix is of the \textit{difference type}, the Yang-Baxter equation imposes the constraint that its dependency on time is linear. 
\begin{align}
f(t)=\alpha t+\beta,
\label{constraintf(t)}
\end{align}
where $\alpha$ and $\beta$ are constants. In addition to this, the S-matrices corresponding to the exchange of the particles with the same chirality also satisfy the Yang-Baxter equation. For three right moving particles, we have
\begin{align}\nonumber
S^{ij}(z_i,z_j)S^{ik}(z_i,z_k)S^{jk}(z_j,z_k)=\\S^{jk}(z_j,z_k)S^{ik}(z_i,z_k)S^{ij}(z_i,z_j).\label{YBn3}\end{align}
Similar expression exists for three left moving particles, which can be obtained by applying the transformation $z_{i,j,k}\rightarrow\bar{z}_{i,j,k}$ to the above equation (\ref{YBn3}).  The Eq~(\ref{constraintf(t)}), together with the constancy of $u$ (\ref{constraintu(t)}), are necessary and sufficient to satisfy all the Yang-Baxter equations (\ref{YBn1}), (\ref{YBn3}), thus resulting in a consistent wavefunction.  Using the relations (\ref{rel1N}), (\ref{rel3N}) and the Yang-Baxter equations (\ref{YBn1}), (\ref{YBn3}), one can express any amplitude in the N-particle wavefunction in terms of any one amplitude. Similar to the one particle case, this amplitude can be determined by applying periodic boundary conditions. We may choose this to be $f^{N...1,\{\chi_j\}}_{\sigma_1...\sigma_N}(\bar{z}_1,...,z_N)$. Applying periodic boundary conditions yields the following relation
\begin{align}
f^{j...,\{\chi_i\}}(\bar{z}_1,..,z_j,..,z_N)=f^{...j,\{\chi_i\}} (\bar{z}_1,..,z_j+L,..,z_N).\label{nppbc}\end{align}
Here $j$ is a right moving particle. Similar expression exists for a left moving particle, which can be obtained by applying the transformation $z_j\rightarrow\bar{z}_j$ to the above equation (\ref{nppbc}). Let us consider the transport operator $Z_j$ which transports the particle $j$ around the system once, and without loss of generality we may choose it to be a right mover. Using (\ref{nppbc}), we have

\begin{align}\nonumber
f^{N...1,\{\chi_i\}}(\bar{z}_1,..,z_j-L,..,z_N)=Z_{j}(\bar{z}_1,...,z_N) \\f^{N...1,\{\chi_i\}}(\bar{z}_1,..,z_j,..,z_N).\label{diffeq1}\end{align}
Now, let us construct the transport operator $Z_j$. Starting from the chosen reference amplitude, as we move the particle $j$ to the right, it first encounters the particle $j-1$, which is also a right moving particle. Thus the S-matrix associated with the exchange of these two particles is $S_{jj-1}(z_{j},z_{j-1})$. Recall that the argument of the S-matrix associated with the particles with the same chirality is proportional to the distance between the particles. The particle $j$ then can be moved past all the right movers $j-2,...,N_L+1$ with the corresponding S-matrices $S_{jj-2}(z_{j},z_{j-2}),...,S_{jN_L+1}(z_j, z_{N_L+1})$ respectively. As the particle $j$ is moved further to the right, it encounters particle $N_L$, which is a left mover. The S-matrix associated with this interaction is $S_{jN_L}(z_j,\bar{z}_{N_L})$. Recall that the argument of the S-matrix associated with the scattering of two particles with opposite chiralities is just the time at which the scattering occurs. Similarly, the particle $j$ can be moved past all the left movers $N_L-1,...,1$ with the corresponding S-matrices $S_{jN_{L}-1}(z_j, \bar{z}_{N_L-1}),...,S_{j1}(z_j, \bar{z}_1)$. As the particle $j$ is further moved to the right it then encounters particle $N$, which is a right mover. The S-matrix associated with the exchange of these particles is $S_{jN}(z_j,z_N+L)$. Note that the particle $j$ had to travel the distance $L-x_j+x_N$ to encounter particle $N$, hence the presence of the `shift' $L$. The particle $j$ can be further moved to the right past all the remaining right movers $N-1,...,j+1$ with the corresponding S-matrices $S_{jN-1}(z_j,z_{N-1}+L),...,S_{jj+1}(z_k,z_{j+1}+L)$. Note that the above process can be interpreted as transporting the particle in time. Ignoring the overall scalar factor associated with the S-matrices, the transport operator $Z_{j}(z_1,...,\bar{z}_N)$ which transports the particle $j$ around the system once takes the following form  

  \begin{align}\nonumber
      &Z_j(\bar{z}_1,...,z_N)=S^{jj+1}(z_j,z_{j+1}+L)\dots S^{jN}(z_j,z_N+L)\\ &S^{j1}(z_j,\bar{z}_1)\dots S^{jN_L}(z_j,\bar{z}_{N_L}) \dots S^{jj-1}(z_j,z_{j-1}).\label{diffeq2}
  \end{align}  
Using the exact form of the S-matrices (\ref{2psmat}) along with the relation (\ref{constraintf(t)}) and ignoring the scalar scattering phase, the equations (\ref{diffeq1}) and (\ref{diffeq2}) can be expressed as
\begin{align}\nonumber
f^{N...1,\{\chi_i\}}(\bar{w}_1,..,w_j-\kappa,..,w_N)=Z_{j}(\bar{w}_1,...,w_N) \\f^{N...1,\{\chi_i\}}(\bar{w}_1,..,w_j,..,w_N),\label{diffeq1w}\end{align}
and \begin{align}\nonumber
      &Z_j(\bar{w}_1,...,w_N)=R_{jj+1}(w_{j+1}-w_j+\kappa)\dots\\\nonumber&R_{jN}(w_N-w_j+\kappa) R_{j1}(\bar{w}_1-w_j)\dots R_{jN_L}(\bar{w}_{N_L}-w_j)\\& R_{jN_L+1}(w_{N_L+1}-w_j)\dots R_{jj-1}(w_{j-1}-w_j),\label{diffeq2w}
  \end{align}  
respectively. Here we have introduced the following notation
\begin{align} w_j=\alpha z_j, \;\; \;\;\bar{w}_k=\alpha \bar{z}_k, \;\; \;\; \kappa=\alpha L. 
\end{align}
The transport operators satisfy the following relations
\begin{align}\nonumber
  Z_j(\bar{w}_1,...,w_k-\kappa,...,w_N) Z_k(\bar{w}_1,...,w_N)=\\Z_{k}(\bar{w}_1,...,w_j-\kappa,...,w_N)Z_j(\bar{w}_1,...,w_N). \label{transportop}
\end{align}
In the above equation (\ref{diffeq2}), we have chosen $j$ and $k$ to be right moving particles. In the case of left moving particles, we simply need to apply the transformation $z_{j/k}\rightarrow\bar{z}_{j/k}$ in the above equation along with the change in the sign of $\kappa$. Eq.~(\ref{transportop}) is the \textit{discrete zero-curvature condition}. It guarantees that transporting particle $j$ and then particle $k$ gives the same result as performing the two operations in the opposite order. The Yang-Baxter equation ensures this compatibility provided $u$ is constant and $f(t)$ is linear in time. The equations (\ref{diffeq1w}) and (\ref{diffeq2w}) are nothing but the quantum Knizhnik-Zamolodchikov (qKZ) equations associated with the XXZ R-matrix. The solution of the qKZ equations therefore determines the reference amplitude, while all remaining amplitudes corresponding to different orderings follow by successive application of the two-particle scattering matrices. The development of the qKZ equations emerged from early studies of conformal field theory and integrable field theories. The Knizhnik-Zamolodchikov (KZ) equations were introduced by Knizhnik and Zamolodchikov \cite{KnizhnikZamolodchikov} as differential equations governing correlation functions in conformal field theories with affine Lie algebra symmetry. A quantum deformation of the KZ equations (qKZ) first appeared in the work of Smirnov \cite{Smirnov}, where such equations arose as fundamental relations satisfied by form factors in the sine-Gordon model. They were subsequently derived within the representation theory of quantum affine algebras by Frenkel and Reshetikhin \cite{Frenkel}. Explicit solutions in terms of Jackson integrals and off-shell Bethe vectors were first constructed for the quantum affine algebra $\mathfrak{sl}_2$ by Reshetikhin \cite{rishetikhin1,rishetikhin2} and later generalized to $\mathfrak{sl}_n$ by Tarasov and Varchenko \cite{Tarasov:1993vs,Tarasov:1994bb,varchenko}. The connection between solutions of the qKZ equations and the off-shell Bethe ansatz was subsequently formulated systematically by Babujian et al. \cite{Babujian_1997}. Thus, using the methods developed in the above mentioned works, one can solve the qKZ equations (\ref{diffeq1}), (\ref{diffeq2}) to obtain the exact wavefunction. It was recently show in \cite{pasnoorigrossneveu2}, that the solutions to the qKZ equations exhibit a variational structure involving Yang-Yang action which enables one to extract exact long time dynamics.

To summarize, we have constructed the $N$-particle wavefunction (\ref{npwf1}), (\ref{npwf2}) that satisfies the time-dependent Schrodinger equation (\ref{SE}). This construction gives rise to the two-particle $S$-matrices (\ref{2psmat}), (\ref{XXZRmat}) which, unlike those appearing in the standard Bethe ansatz, depend explicitly on the spacetime coordinates of the particles. These $S$-matrices satisfy the quantum Yang-Baxter equations (\ref{YBn1}), (\ref{YBn3}) ensuring the consistency of the many-body wavefunction and allowing all amplitudes to be related to a single reference amplitude. To determine the wavefunction explicitly, we construct a transport operator (\ref{diffeq2}) from the two-particle $S$-matrices that transports a particle through the system. Imposing periodic boundary conditions then leads to a set of matrix difference equations (\ref{diffeq1}), namely the qKZ equations, which impose constraints on the reference amplitude. Together, these results constitute the generalized Bethe-ansatz solution of the time-dependent $U(1)$-symmetric Thirring model.

\subsection{Constraints on integrability and the renormalization-group flow}

We now show that the constraints imposed by the generalized Bethe ansatz reproduce the RG flow of the corresponding static model. Since the relation between the bare couplings and the spectral and crossing parameters is in general non universal, we restrict this comparison to the universal weak-coupling regime,
\begin{equation}
|g_{\parallel}|,|g_{\perp}|\ll 1.
\end{equation}
Expanding Eq.~(\ref{u(t)def}) to quadratic order gives

\begin{equation}
u^2=g_{\parallel}^2(t)-g_{\perp}^2(t)\equiv Q.\label{Qsmall}
\end{equation}
Because $u$ is independent of time, the quantity $Q$ is a dynamical invariant. This is precisely the RG invariant of the anisotropic current-current perturbed $SU(2)_1$ WZW model (\ref{WZWccRGinv}). In the same weak-coupling regime, Eq.~(\ref{f(t)def}) becomes
\begin{equation}
f(t)=\operatorname{arccoth}\left(\frac{g_{\parallel}(t)}{\sqrt{Q}}\right).
\label{f(t)small}
\end{equation}
Using the integrability condition (\ref{constraintf(t)}) in (\ref{f(t)small}) and differentiating it with respect to time we obtain
\begin{equation}
\frac{d g_{\parallel}}{dt}=-\frac{\alpha}{\sqrt{Q}}g_{\perp}^2.
\label{gparallel_timeflow}
\end{equation}
Differentiating (\ref{Qsmall}) and using the above relation, one obtains
\begin{equation}
\frac{d g_{\perp}}{dt}=-\frac{\alpha}{\sqrt{Q}}
g_{\parallel}g_{\perp}.
\label{gperp_timeflow}
\end{equation}
Thus the generalized Bethe ansatz does not permit arbitrary functions $g_{\parallel}(t)$ and $g_{\perp}(t)$. Quantum integrability restricts their temporal evolution to the coupled equations (\ref{gparallel_timeflow}) and (\ref{gperp_timeflow}). These equations take the same form as the RG equations (\ref{RGeq}) of the anisotropic current-current perturbed $SU(2)_1$ WZW model, provided we identify the time $t$ with the logarithm of the energy scale $\ell$:

\begin{equation}
\ell=\frac{\alpha}{4\sqrt{Q}}t+\ell_0.
\label{RG_time}
\end{equation}
Here $\ell_0$ is an arbitrary additive constant. 
In the static theory, $\ell$ denotes the logarithm of an energy or length scale. In the generalized Bethe-ansatz solution, the same trajectory arises from the consistency of the time-dependent many-body wavefunction, and $\ell$ is related linearly to the physical time through Eq.~(\ref{RG_time}). The RG trajectory is therefore dynamically realized by the temporal evolution of the Hamiltonian. We emphasize that this correspondence has not been imposed by choosing the couplings to satisfy the RG equations. We began with arbitrary time-dependent interaction strengths $g_{\parallel}(t)$ and $g_{\perp}(t)$. The two-particle Schrodinger equation determines the corresponding XXZ scattering matrix. Consistency of the many-body wavefunction then requires a time-independent crossing parameter $u$ and a spectral parameter $f(t)$ that is linear in time. In the universal weak-coupling regime, the first is identified with the constancy of the parameter $Q$ along the RG flow, that is it becomes the RG invariant. The second one corresponds to the RG flow of the coupling strengths through the parameter $f$ (\ref{paramf}) of the static model \cite{pasnoorigrossneveu2}. The RG flow of the static theory therefore emerges as the condition for quantum integrability of the time-dependent theory.


\section{RG limit cycles $\equiv$ periodic time evolution}
\label{sec:periodic_driving}
In the previous section we have constructed the generalized Bethe ansatz wavefunction and we have shown that the constraints imposed by integrability are exactly same as the RG equations of the static model. In this section we resolve these constraints and obtain the explicit expressions of the time-dependent strengths. We then analyze the temporal periodicity of the integrable interaction strengths in the cyclic regime. We first consider the universal weak-coupling limit of the generalized Bethe-ansatz solution and compare its period with the one-loop RG flow obtained in Ref.~\cite{LeClair}. We then return to the exact constraints imposed by integrability on the parameters $f(t)$ and $u$, and by using the relations between these parameters and the coupling strengths we show that the resulting exact time-dependent Hamiltonian has twice the weak-coupling period, in agreement with the period obtained from the proposed all-orders RG equations of Ref.~\cite{LeClair}.

Recall that integrability requires that spectral parameter $f(t)$ (\ref{constraintf(t)}), (\ref{f(t)def}) is linear in time while the crossing parameter $u$ is independent of time (\ref{constraintu(t)}). In the cross over regime the relevant parameters $f(t)$ and $u$ are purely imaginary. Thus it is useful to define $\bar{f}(t)=-if(t)$, $\bar{u}=-iu$, such that 
\begin{equation}
\bar{f}(t)\equiv\alpha_I t+\beta_I,
\label{fbar_cycle}
\end{equation}
where $\alpha=i\alpha_I$, $\beta=i\beta_I$. 

\paragraph{Weak-coupling solution and the one-loop RG period.}
Let us now consider the coupled equations describing the integrable time evolution of the coupling strengths $g_{\parallel}(t)$ and $g_{\perp}(t)$ (\ref{gparallel_timeflow}), (\ref{gperp_timeflow}). These equations can be solved to obtain the explicit expressions of the coupling strengths for small values of the couplings. In the cross-over region $|g_{\perp}|>|g_{\parallel}|$, using the above introduced parametrization, the solutions to these equations are given by 

\begin{equation}
g_{\parallel}(t)
=\frac{h}{4}\cot\bar{f}(t)\qquad g_{\perp}(t)=\frac{h}{4}|\csc\bar{f}(t)|,
\label{weak_GBA_cyclic}
\end{equation}
up to the choice of branch for $g_{\perp}$. In the above expression, we have used the definition of $h$ (\ref{hdef}). Under a shift
\begin{equation}
\bar{f}(t)\rightarrow\bar{f}(t)+\pi,
\end{equation}
the two couplings transform according to
\begin{equation}
g_{\parallel}(\bar{f}(t)+\pi)=g_{\parallel}(\bar{f}(t)),\qquad g_{\perp}(\bar{f}(t)+\pi)=g_{\perp}(\bar{f}(t)).
\label{weak_pi_shift}
\end{equation}
Hence the corresponding period in physical time is
\begin{equation}
T_{\rm weak}=\frac{\pi}{|\alpha_I|}.
\label{weak_physical_period}
\end{equation}
To compare the above physical-time period with the 1-loop RG period (\ref{LeClair1loop}) obtained in Ref.~\cite{LeClair}, consider the relation between time and the logarithm of the cutoff (\ref{RG_time}). Using the parametrization in the cross-over regime, we have

\begin{align}\ell-\ell_0=\frac{|\alpha_l|}{h}t. \label{elly}
\end{align}
Using the physical-time period in the above equation, we find that one weak-coupling temporal cycle corresponds to the following interval in RG time
\begin{equation}
\Delta\ell_{\rm weak}=\frac{|\alpha_I|}{h}T_{\rm weak}=\frac{\pi}{h}.\label{weak_RG_period}
\end{equation}
This is precisely the one-loop period (\ref{LeClair1loop}) obtained in Ref.~\cite{LeClair}. Thus the universal weak-coupling solution of the generalized Bethe ansatz reproduces both the one-loop BKT trajectory and its RG period. The weak-coupling expressions in Eq.~(\ref{weak_GBA_cyclic}), however, diverge whenever
\begin{equation}
\theta=n\pi,
\qquad n\in\mathbb Z.
\end{equation}
indicating that the one-loop solution obtained above cannot describe the evolution between successive strong-coupling singularities where the RG trajectory restarts on the next branch: When $g_{\parallel}=0$, the value of $g_{\perp}$ takes its minimum value. As $g_{\parallel}$ is increased, the value of $g_{\perp}$ simultaneously increases. As $g_{\parallel}(t)\rightarrow\infty$, the strength $g_{\perp}\rightarrow\infty$, after which the cycle restarts at $g_{\parallel}\rightarrow-\infty$ with $g_{\perp}\rightarrow\infty$. As $g_{\parallel}$ increasess from $-\infty$ towards zero, the value of $g_{\perp}$ decreases reaching its minimum when $g_{\parallel}=0$, and the cycle repeats indefinitely. Consequently, the perturbative solution does not determine how the trajectory should be continued globally through these strong-coupling regions where the coupling strengths diverge. This information is contained in the exact generalized Bethe-ansatz relations which we shall discuss in the following.

\paragraph{Exact coupling strengths and the complete cycle.}
The exact expressions of the time-dependent coupling strengths can be obtained by using the integrability condition (\ref{constraintf(t)}) in the equations (\ref{f(t)def}) and (\ref{u(t)def}) which relate the coupling strengths with the parameters $f(t)$ and $u$. Using the parametrization of the cross-over regime, one obtains

\begin{equation}
g_{\perp}(t)=\arccos\left[\frac{\sin(\alpha_I t+\beta_I)}{\sqrt{\sinh^2\bar{u}+\sin^2(\alpha_I t+\beta_I)}}\right],
\label{gperp_exact_final}
\end{equation}
and
\begin{equation}
g_{\parallel}(t)=\arccos\left[\frac{\cosh\bar{u}\sin(\alpha_I t+\beta_I)}{\sqrt{\sinh^2\bar{u}+\sin^2(\alpha_I t+\beta_I)}}\right].
\label{gparallel_exact_final}
\end{equation}
Now consider the shift $\bar{f}(t)\rightarrow\bar{f}(t)+\pi$
\begin{equation}
g_{\perp}(\bar{f}(t)+\pi)=\pi-g_{\perp}(\bar{f}(t)),
\end{equation}
and
\begin{equation}
g_{\parallel}(\bar{f}(t)+\pi)=\pi-g_{\parallel}(\bar{f}(t)).
\label{exact_half_cycle_transform}
\end{equation}
Thus an advance of $\pi$ in the spectral phase does not return the exact Hamiltonian to its initial value. Only after a second shift,
\begin{equation}
\bar{f}(t)\rightarrow\bar{f}(t)+2\pi,
\end{equation}
do both interaction strengths return:
\begin{equation}
g_{\parallel}(\bar{f}(t)+2\pi)=g_{\parallel}(\theta),\qquad g_{\perp}(\bar{f}(t)+2\pi)=g_{\perp}(\bar{f}(t)).
\end{equation}
The exact period in physical time is therefore
\begin{equation}
T_{\rm exact}= \frac{2\pi}{|\alpha_I|}=2T_{\rm weak}.
\label{Texact_final}
\end{equation}
The factor of two does not arise from a different time dependence of the spectral parameter as both the weak-coupling and exact solutions arise from the same integrability constraint (\ref{constraintf(t)}). Rather, it results from the global continuation through the strong-coupling region. In the one-loop solution the trajectory reaches infinity after an advance $\Delta\bar{f}(t)=\pi$ and is identified with the beginning of the next RG cycle. The exact generalized Bethe-ansatz solution resolves this region at finite coupling. After the same advance $\Delta\bar{f}(t)=\pi$, it reaches the distinct point
\begin{equation}(g_{\parallel},g_{\perp})\longrightarrow(\pi-g_{\parallel},\pi-g_{\perp}),
\end{equation}
and a second interval of length $\pi/|\alpha_I|$ is required to return to the original Hamiltonian. Finally, using the normalization between physical and RG time fixed in the weak-coupling region (\ref{elly}), the exact temporal period corresponds to
\begin{equation}
\Delta\ell_{\rm exact}= \frac{|\alpha_I|}{h}T_{\rm exact}=\frac{2\pi}{h}.
\end{equation}
Thus
\begin{equation}
\Delta\ell_{\rm exact}=\frac{2\pi}{h}.
\label{GBA_exact_RG_period_final}
\end{equation}
This is precisely the period (\ref{LeClairallloop}) obtained in Ref.~\cite{LeClair} from their proposed all-orders beta function, which is twice the one-loop result (\ref{LeClair1loop}). We therefore obtain the correspondence
\begin{equation}
T_{\rm weak}=\frac{\pi}{|\alpha_I|}\quad\longleftrightarrow\quad\lambda_{\rm 1-loop}=\frac{\pi}{h},
\end{equation}
while the exact generalized Bethe-ansatz evolution gives
\begin{equation}
T_{\rm exact}=\frac{2\pi}{|\alpha_I|}\quad\longleftrightarrow\quad\lambda_{\rm all-orders}=\frac{2\pi}{h}.
\end{equation}
Thus, the generalized Bethe ansatz reproduces the one-loop cyclic period in its universal weak-coupling limit, while its exact solution exhibits the doubled period characteristic of the all-loop cyclic RG period. Note that, the detailed form of the bare couplings away from weak coupling is regularization dependent, and therefore the functional forms of the exact generalized Bethe-ansatz couplings in Eqs.~(\ref{gparallel_exact_final}) and (\ref{gperp_exact_final}) do not coincide with the functional forms of the solutions of the proposed all-orders beta functions of Ref.~\cite{LeClair,Gerganov}. The period of a limit cycle, however, is a regularization independent property of the RG flow. It is therefore significant that the complete generalized Bethe-ansatz evolution doubles the weak-coupling period in exactly the same manner as the proposed all-orders RG flow.

Finally, we note that in the cyclic C regime considered above, where $ |g_{\parallel}(t)|<|g_{\perp}(t)|$, the XXZ $R$-matrix associated with the two particle S-matrices is expressed in terms of trigonometric functions. This ultimately leads to the periodic time dependence of the integrability-preserving couplings. By contrast, for $ |g_{\parallel}(t)|>|g_{\perp}(t)| $, corresponding to the AF regime for $g_{\parallel}(t)>0$ and the WC regime for $g_{\parallel}(t)<0$, the XXZ $R$-matrix is expressed in terms of hyperbolic functions (\ref{XXZRmat}). Due to which, the integrability preserving time-dependent coupling strengths are monotonic functions \cite{PasnooriRG}. In the universal weak-coupling limit, their temporal trajectories coincide with the corresponding RG flows in the AF and WC regimes. Unlike the cyclic C regime, these monotonic flows terminate at, or originate from, the corresponding RG fixed points. 

\section{Conclusion and outlook}
\label{sec:conclusion}

In this work, we have established a direct connection between cyclic renormalization-group flows and the real-time dynamics of an integrable quantum field theory. We considered the anisotropic current-current perturbed $SU(2)_1$ WZW model whose static renormalization-group flow exhibits cyclic trajectories. Through non-Abelian bosonization this model is equivalent to the interacting isospin sector of the $U(1)$-symmetric Thirring model. We considered this model with explicitly time-dependent longitudinal and transverse interaction strengths, $g_{\parallel}(t)$ and $g_{\perp}(t)$ and solved it using the generalized Bethe ansatz framework. The two-particle Schrodinger equation determines the corresponding XXZ scattering matrices, while consistency of the many-body wavefunction and periodic boundary conditions reduce the problem to quantum Knizhnik-Zamolodchikov equations associated with the XXZ $R$-matrix.

The quantum integrability severely constrains the allowed time dependence of the interaction strengths such that the crossing parameter $u$ of the XXZ $R$-matrix is required to remain constant, while the corresponding spectral parameter $f(t)$ evolves linearly with physical time. In the universal weak-coupling regime, upon an appropriate linear identification of physical time with the logarithmic RG scale, these conditions on the interaction strengths precisely become the one-loop RG equations of the anisotropic current-current perturbed $SU(2)_1$ WZW model. Thus, the correspondence between time evolution and RG flow is not imposed on the Hamiltonian as an external driving protocol; rather, it emerges as a consequence of the consistency conditions required for quantum integrability.

The result becomes particularly interesting in the crossover regime $Q<0$, where the static RG trajectories are cyclic. Under the correspondence established here, the RG cycle becomes a cyclic evolution of the interaction strengths in physical time. In this sense, the RG limit cycle acquires a dynamical realization: translation along the logarithmic RG scale is replaced by translation in physical time. The discrete scale invariance associated with the RG limit cycle is consequently mapped onto a discrete time-translation structure of the driven quantum system. In particular, we find that the periodicities obtained from the generalized Bethe-ansatz solution are in direct agreement with the cyclic RG analysis of \cite{LeClair}. In the universal weak-coupling limit, the time-dependent interaction strengths repeat after a physical-time interval $T_{\rm weak}=\pi/|\alpha_I|$, which, using the relation between physical and RG time, corresponds to an RG period $\lambda_{\rm 1-loop}=\pi/h$, precisely matching the one-loop result of \cite{LeClair}. By contrast, the exact generalized Bethe-ansatz solution allows one to reach the strong-coupling region and requires  $T_{\rm exact}=2\pi/|\alpha_I|$ before the Hamiltonian returns to its initial form, and hence resulting in $\lambda_{\rm all-orders}=2\pi/h$. This coincides with the period obtained from the proposed all-orders RG flow of \cite{LeClair}, demonstrating that the factor of two enhancement of the cycle period beyond one loop is reproduced exactly by the generalized Bethe-ansatz solution.  In this sense, the correspondence between RG scale and physical time is more than a formal analogy: an RG limit cycle is converted into an actual periodic evolution of an interacting quantum field theory whose many-body dynamics remain exactly tractable.

The connection established here between cyclic RG flow and periodic real-time evolution suggests that phenomena usually formulated in coupling space may acquire direct dynamical realizations in periodically driven integrable systems. The standard Bethe ansatz has been very successful in probing novel phases such as symmetry protected topological  (SPT) phases \cite{PAA1,PAA2,SUSYpaper,pasnooriduality25,PasnooriSGPT}, symmetry breaking phases and non perturbative effects associated with impurities \cite{Parmeshkondo1,Parmeshkondo2,PasnooriXXZPD,NHKPRB,PTXXZPRB,XXZKondo,XXXKondo}. Thus, extending the present generalized Bethe ansatz construction to open boundaries would therefore
be particularly interesting, since Floquet systems can support intrinsically dynamical phases with no equilibrium counterpart. Such a framework generalizing the time-dependent integrability to incorporate boundaries was recently formulated in \cite{pasnooriBGBA}. In this setting, the bulk XXZ scattering matrices are supplemented by boundary reflection matrices and the many-body wavefunction is governed by the boundary qKZ equations. The relevant object in this framework is the transport operator which corresponds to a complete evolution over one cycle rather than any individual instantaneous Hamiltonian. This raises the possibility that the qKZ transport operators or the corresponding monodromy may encode dynamical topological information, potentially providing an exact route to investigating the emergence of Floquet symmetry-protected topological phases which are intrinsically dynamical and having no analogue in a static Hamiltonian \cite{YaoFSPT,NayakFSPT,RoyFSPT}.

The exact solution also provides a controlled setting to study spontaneous breaking of discrete time-translation symmetry. One may ask whether physical observables evaluated in the exact time-dependent state develop a robust subharmonic response giving rise to Floquet time crystal \cite{TCNayak,KhemaniTC}. A closely related question concerns Floquet heating. Generic interacting periodically driven systems tend to absorb energy \cite{Rubio}, whereas the present drive is constrained by quantum integrability. It would therefore be interesting to determine whether the infinite set of integrability preserving constraints suppresses unrestricted heating or produces a distinct non-perturbative mechanism for stabilizing long-lived dynamical phases. In addition to Hermitian systems discussed above, periodically driven non Hermitian systems are shown to exhibit new phenomena such as Floquet dissipative quasicrystals \cite{NHFloquetWu,NHFloquetWeidemann}. It was recently shown using generalized Bethe ansatz that in non-Hermitian systems, the set of integrability preserving time-dependent strengths is larger than the set corresponding to the RG flows \cite{pasnooriNHKtime}. Thus, integrability preserving periodic driving in non-Hermitian systems is expected to host a rich variety of novel phenomena. 

 Finally, these ideas may be amenable to experimental realization \cite{ZhangSFSPT,DTCgoogle}. The factorized structure of the evolution in terms of XXZ $R$-matrices makes the construction naturally suited to `digital quantum circuits' \cite{Prozen,Vanicat,EVtrotter,EVcircuitSM, Quantumcircuitdynamics}, where the corresponding two-body scattering matrices can be implemented as quantum gates, albeit care should be taken in simulating integrable dynamics with time-dependent parameters. Similarly, `analog quantum-simulation platforms' including cold-atom experiments \cite{coldatomsBA,zollerkondocold} and superconducting-circuit architectures \cite{KochpotSC,circuitSPT,roy1,IoffeSC,KuzminSC,LarkinSC}, offer possible routes \cite{pasnoori2026sgcircuit} to engineering the required time-dependent anisotropic interactions directly. Simulation of time-dependent dynamics on digital and analog superconducting circuits is the subject of our forthcoming work. 

\bibliography{refpaper}

\end{document}